\documentclass[conference]{IEEEtran}
\IEEEoverridecommandlockouts
\usepackage{cite}
\usepackage{amsmath,amssymb,amsfonts}
\usepackage{algorithmic}
\usepackage[pdftex]{graphicx}
\usepackage{textcomp}
\usepackage{xcolor}
\usepackage{booktabs}
\usepackage{subcaption}

\usepackage{tikz}
\usetikzlibrary{shapes.geometric, arrows, calc, positioning}
\tikzstyle{startstop} = [rectangle, rounded corners, minimum width=3cm, minimum height=1cm,text centered, draw=black, fill=red!10]
\tikzstyle{process} = [rectangle, minimum width=3.5cm, minimum height=1cm, text centered, draw=black, fill=blue!10]
\tikzstyle{decision} = [diamond, minimum width=3cm, minimum height=1cm, text centered, draw=black, fill=green!10, aspect=2]
\tikzstyle{arrow} = [thick,->,>=stealth]

\def\BibTeX{{\rm B\kern-.05em{\sc i\kern-.025em b}\kern-.08em
    T\kern-.1667em\lower.7ex\hbox{E}\kern-.125emX}}
\begin{document}

\title{Planning Autonomous Vehicle Maneuvering in Work Zones Through Game-Theoretic Trajectory Generation \\}

\author{\IEEEauthorblockN{Mayar Nour}
\IEEEauthorblockA{\textit{Civil and Environmental Engineering} \\
\textit{University of Western Ontario}\\
London, Ontario, Canada \\
mnour6@uwo.ca}
\and
\IEEEauthorblockN{Atrisha Sarkar}
\IEEEauthorblockA{\textit{Electrical and Computer Engineering} \\
\textit{University of Western Ontario}\\
London, Ontario, Canada \\
atrisha.sarkar@uwo.ca}
\and
\IEEEauthorblockN{Mohamed H. Zaki}
\IEEEauthorblockA{\textit{Civil and Environmental Engineering} \\
\textit{University of Western Ontario}\\
London, Ontario, Canada \\
m.zaki@uwo.ca}
}

\maketitle

\begin{abstract}
Work zone navigation remains one of the most challenging manoeuvres for autonomous vehicles (AVs), where constrained geometries and unpredictable traffic patterns create a high-risk environment. Despite extensive research on AV trajectory planning, few studies address the decision-making required to navigate work zones safely. This paper proposes a novel game-theoretic framework for trajectory generation and control to enhance the safety of lane changes in a work zone environment. By modelling the lane change manoeuvre as a non-cooperative game between vehicles, we use a game-theoretic planner to generate trajectories that balance safety, progress, and traffic stability. The simulation results show that the proposed game-theoretic model reduces the frequency of conflicts by 35 percent and decreases the probability of high risk safety events compared to traditional vehicle behaviour planning models in safety-critical highway work-zone scenarios.
\end{abstract}


\section{Introduction}
Work zones represent one of the most safety-critical segments of highway infrastructure, because they involve temporary lane closures, geometric constraints, and introduce disruptive traffic flow patterns. These characteristics create complex interaction challenges during lane changes, which are particularly difficult for Autonomous Vehicles (AVs)\cite{maddineni2024safety}. In traffic flow, lane changes are categorised as discretionary or mandatory. Discretionary lane changes are made to improve driving comfort or speed. On the other hand, mandatory lane changes are a must to follow a route or obey road rules \cite{zheng2014recent}. In work zones, due to lane closures, lane changes become a mandatory manoeuvre. Since these manoeuvres are imposed by the road environment rather than initiated voluntarily by the driver, they can increase cognitive load and time pressure. This usually increases the likelihood of rear-end conflicts, and unsafe gap acceptance behaviour \cite{bidkar2023effect}\cite{rangaswamy2024analysis}. Despite advancements in traffic management systems, work zone crashes remain higher than those on regular highway segments \cite{ansari2025systematic}. According to the Federal Highway Administration (FHWA) and the National Safety Council, annual fatalities in work zones  reached 898 in recent years, representing an increase of nearly 50\% in the last decade. In particular, more than 60\% of these fatalities are vehicle drivers, highlighting the risks of vehicle interactions in work zone environments \cite{art_nsc_stats}\cite{fhwa_workzone_stats}. Due to the interactive nature of the navigation task, vehicles must reason strategically in work zones. A strategic interaction occurs when the decision of one vehicle directly affects, and is affected by, the decisions of other vehicles. This type of interaction is necessary during mandatory lane changes, where drivers must negotiate gaps and adjust speeds accordingly, depending on the actions of other vehicles. AVs offer the potential to mitigate these risks by leveraging advanced strategic trajectory planning techniques. Game theory (GT) is a mathematical framework for modelling strategic interactions between agents \cite{arbis2019game}. For automated driving, by treating interacting vehicles as rational agents with competing objectives, such as, safety and efficiency, the lane change decision can be formulated as a non-cooperative game, \cite{ji2020review} offering a structured mechanism for resolving conflicts and generating coordinated trajectories. Although prior studies have investigated general game-theoretic trajectory planning for AVs \cite{yan2023multi} \cite{cai2024game}, the focus has been limited to safety-critical scenarios, such as work zones.

This paper proposes a game-theoretic framework for trajectory generation and control to improve lane change safety in work zone environments. The proposed model identifies interacting vehicle pairs upstream of a lane closure and generates optimal trajectories for each vehicle based on a multi-objective utility function. Lane change decisions must balance multiple objectives rather than optimise a single factor, as work zones involve safety constraints, traffic flow efficiency, and compliance with road regulations simultaneously. Therefore, the utility function accounts for collision risk, progress toward the target lane, and traffic regulation.

Since work zones increase the risk of unsafe lane change interactions, the proposed framework is evaluated from a safety perspective. To evaluate the safety implications of the proposed approach, we compare the game-theoretic lane change framework with a baseline of standard car-following and lane-changing models without strategic interaction.
Each scenario is executed in multiple random seeds to ensure statistical robustness. Safety performance is quantified using safety surrogate measures, specifically Time-To-Collision (TTC), alongside conflict frequency analysis across simulation time.

The contributions of this work are as follows:
\begin{itemize}
\item A quantitative assessment of the dynamic non-cooperative game theory model on work zone lane change safety using safety surrogate measures.
    \item The design of a multi-objective utility function that balances safety, efficiency, and traffic regulation for work zone merges.
    \item Integration of a game-theoretic lane change trajectory planner within a microscopic traffic simulator to enable lane change manoeuvres within a work zone.
    
\end{itemize}

The remainder of the paper is organised as follows. Section II reviews the related work. Section III presents the proposed methodology, including the simulation setup, vehicle modelling, and game-theoretic framework. Section IV presents the results and the evaluation metrics. Finally, Section V concludes the paper and outlines limitations and potential future work.
\section{Related Work}

This section reviews previous research on work zone safety and the impact of AVs in such environments. It defines trajectory planning for AVs and game-theoretic decision-making. Together, these studies identify the gap addressed in this work.

\subsection{Work Zone Safety and AVs Impact}
Work zones are associated with higher crash and conflict rates compared to normal motorway segments \cite{theofilatos2017meta}. With rear-end and multi-vehicle collisions being the most common types, as confirmed in \cite{ansari2025systematic}. In addition, traffic volume has been identified as a significant factor in injury crashes. As demand increases, the likelihood of a crash increases. Recent research has investigated the impact of AVs within work zones. Simulation-based studies suggest that different AV control strategies can improve traffic stability and merging efficiency \cite{maddineni2024safety}.  However, work zones increase cognitive demand and delay response times in partially automated systems, highlighting the importance of automation level in safety outcomes \cite{nagy2024evaluation}. Cooperative lane change strategies based on state-of-the-art models, such as Minimising Overall Braking Induced by Lane Changes (MOBIL) and Intelligent Driver Model (IDM), have been shown to reduce merging conflicts in work zones\cite{hussain2024cooperative}. Despite these efforts, safe and efficient merging in work zones remains a challenge.

\subsection{Trajectory Planning in AVs}

Trajectory planning is a fundamental component of autonomous driving systems, which can be used to improve complex manoeuvres such as lane change for AVs. Trajectory planning aims to generate feasible vehicle motions that satisfy safety, comfort, and traffic efficiency \cite{xia2024survey}. Optimisation-based approaches, particularly Model Predictive Control (MPC), are widely used due to their ability to define kinematic constraints and collision avoidance within a rolling horizon framework. For example, \cite{wang2019trajectory} integrated motion prediction with MPC and evaluated lane change safety using crash probability metrics in intersection scenarios. In work zone scenarios, trajectory planning has been formulated as a mixed-integer linear programming to optimise acceleration and lane choices \cite{ma2022trajectory}. 
Learning-based methods, such as Reinforcement Learning (RL), have also been used for trajectory planning. Hierarchical RL frameworks have been proposed for high-level manoeuvre selection and low-level trajectory generation for lane change \cite{naveed2021trajectory}. However, safety guaranteed trajectory planning for merging interactions in a constrained environment like a work zone remains limited.

\subsection{Game Theory Approaches}
Traditional trajectory planning often treats surrounding vehicles as moving obstacles with fixed or independently predicted trajectories. However, such approaches neglect that each vehicle's decision influences and is influenced by others. Ignoring this interdependence may lead to overly conservative or unsafe behaviours in competitive manoeuvres such as merging and overtaking \cite{fisac2019hierarchical}.

Game-theoretic frameworks address this limitation by modelling vehicles as rational agents that optimise their objective while accounting for the responses of others. In overtaking and merging scenarios, equilibrium-based approaches have shown improved safety by capturing the strategic interaction between vehicles \cite{cai2024game}. Moreover, recent work highlights that non-cooperative and Stackelberg game formulations provide structured methods to balance safety and awareness of interactions \cite{yan2023multi}. 
Extensions such as hypergames also consider incomplete or asymmetric information, which allows the modelling of scenarios with occlusions or sensing limitations \cite{kahn2022know}.

These studies indicate that trajectory planning in interactive and constrained environments cannot be treated as a purely geometric or optimisation problem. Instead, it requires explicit modelling of strategic behaviour among agents. 

Despite these advances, game-theoretic trajectory planning in work-zone environments remains under explored. Previous research has independently studied work zone safety, trajectory planning, and game-theoretic decision-making. Limited work has proposed a game-theoretic trajectory planner in a work zone environment. And few studies have evaluated this approach in full traffic simulation using surrogate safety metrics at different automation levels. This study addresses this gap through a co-simulation framework that integrates SUMO and a game-theoretic planner.

\section{Methodology}
This section presents the strategic game-theoretic trajectory generation framework for vehicles navigating a lane closure in a work zone. Vehicles approaching a work zone must carefully adjust their speed and perform a mandatory lane change before the lane closure. The framework defines the two vehicles interacting to perform a lane change upstream of the work zone as the ego and follower vehicle, respectively. The game-theoretic model then represents this interaction as a non-cooperative game between two agents and generates the required trajectories using a Nash equilibrium solver to enable a smooth merge based on a safety utility function. The framework uses a receding horizon control, by sampling the trajectories for a time frame of 6 seconds with a 2-second re-planning frequency to account for the changing state of the environment to ensure a safe lane change. 

\subsection{Simulation Environment and Traffic Model}
\begin{figure}[tbp]
    \centering
    \includegraphics[width=1\columnwidth]{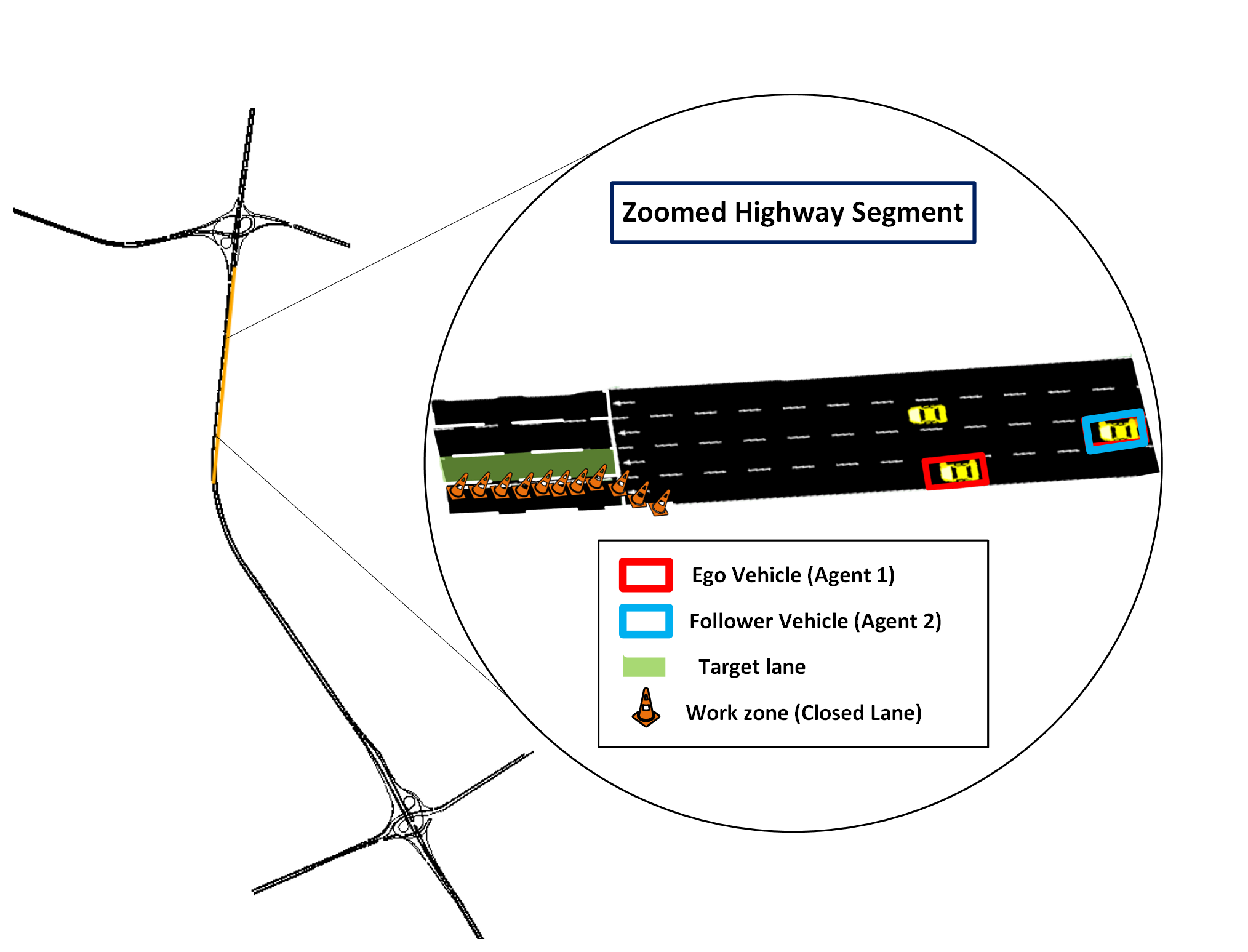}
    \caption{\footnotesize SUMO representation of the 7 km M50 motorway segment in Dublin, Ireland, highlighting the work zone study area.}
    \label{fig:m50_network}
\end{figure}

The proposed approach is evaluated using a microscopic traffic simulation framework based on SUMO (Simulation of Urban MObility)\cite{8569938}. The simulation is conducted on a real-world motorway segment corresponding to the 7 km M50 4-lane motorway in Dublin, Ireland, as shown in Figure~\ref{fig:m50_network}.
The traffic model parameters used in this study are adopted from the calibrated configurations reported in \cite{gueriau2020quantifying}. That work provides calibrated behaviour models based on real traffic data to realistically simulate traffic on the M50 motorway.

The network geometry, lane connectivity, vehicle distribution, and car following model parameters are preserved from the original dataset to ensure realistic traffic dynamics. The setup of the Work zone is introduced by closing the leftmost lane in the southbound direction, with a length of 1 km to simulate a typical work zone scenario. Based on the referenced study, we chose half an hour from 3 pm to 3:30 pm, which represents a saturated traffic load, to evaluate the effectiveness of the model. Moreover, the evaluation considers two levels of automation: Level 2 and Level 4. For each level, the baseline scenario is compared against a corresponding scenario in which vehicles performing lane changes use the proposed game theoretic model. All simulations are executed with a fixed simulation step length of 0.1 s to support fine-grained trajectory control and safety metric extraction. 
Table \ref{tab:simulation_parameters} summarises the parameters used in the simulation.
\begin{table}[tbp]
\caption{Simulation Parameters}
\label{tab:simulation_parameters}
\centering
\begin{tabular}{lll}
\toprule
\textbf{Parameter} & \textbf{Value} \\ 
\midrule
Closed lanes  & 1  \\
Duration & 30 min \\
Time & 15:00-15:30 \\
Step Length & 0.1 sec \\
Trajectory Planning  & 6 sec  \\
Re-planning Frequency  & 2 sec \\
\bottomrule
\end{tabular}
\end{table}

\subsection{Vehicle Modelling and Traffic Dynamics}

In microscopic traffic simulation, vehicle behaviour is governed by two primary sub-models: car-following and lane-changing, which are defined by parameters that can be calibrated to reproduce a specific travel and traffic behaviour \cite{zheng2014recent}. Car-following models regulate longitudinal movement, which determines vehicle acceleration and braking based on the behaviour of the leading vehicle to maintain safe following distances. On the other hand, lane-changing models regulate lateral movement and manage transitions between lanes.

The simulation parameters for both vehicle types used in this study are summarised in Table \ref{tab:parameters}. 
While Avs level 2 are modelled using C-ACC, 
AVs level 4 utilise the Intelligent Driver Model (IDM) following the approach in \cite{gueriau2020quantifying} with a reduced headway $(\tau)$ of 0.6 s compared to 0.8 s and a smaller minimum gap of 1 m compared to 1.5 m. Additionally, Level 4 AVs are configured with higher cooperation levels (1.0) compared to 0.5, which allows drivers to slow down or speed up to create a gap for other vehicles. Both have the same imperfection (sigma $\sigma$), which introduces randomness in driving behaviour, and speed deviation values, which define the standard deviation of the vehicle's speed as a fraction of its desired speed. The values of these parameters reflect the more precise and collaborative nature of level 4 automated driving systems compared to level 2 and human drivers. The behaviour of lane change is modelled using SUMO’s sub-lane (SL2015) lane change model \cite{semrau2016simulation}, which incorporates strategic and cooperative components and allows vehicles to mimic real lane change behaviour by allowing vehicles to be in two different lanes at the same time, instead of the binary behaviour of the default SUMO lane change model. A lateral resolution value partitions the lane into sublanes of that width, allowing vehicles to occupy intermediate positions while performing a lane change. We used a lateral resolution of 0.8 m, which divides the lane into 4 equal sub-lanes; the lower the value, the finer the decision granularity.

\begin{table}[tbp]
\caption{car following and Lane change parameters}
\label{tab:parameters}
\centering
\begin{tabular}{lll}
\toprule
\textbf{Parameters} & \textbf{AV L2} & \textbf{AV L4} \\ 
\midrule
Car Following Model  & C-ACC   & IDM \\
Speed deviation  & 0.05   & 0.05 \\
minGap  & 1.5 & 1 \\
tau $\tau$  & 0.8  & 0.6 \\
sigma $\sigma$  & 0.05  & 0.05 \\
Lane Change Model  & SL2015   & SL2015 \\
Lc Cooperative  & 0.5 & 1 \\
Lateral Resolution & 0.8 & 0.8 \\
\bottomrule
\end{tabular}
\end{table}

During simulations, AV behaviour follows the car-following and lane-changing dynamics unless overridden by the game-theoretic control module during interactive lane change events. This allows externally generated trajectories to be applied directly while preserving the surrounding traffic behaviour.

\subsection{Game-Theoretic Lane Change Framework}
\begin{figure*}[htbp]
    \centering
    \includegraphics[width=0.8\textwidth]{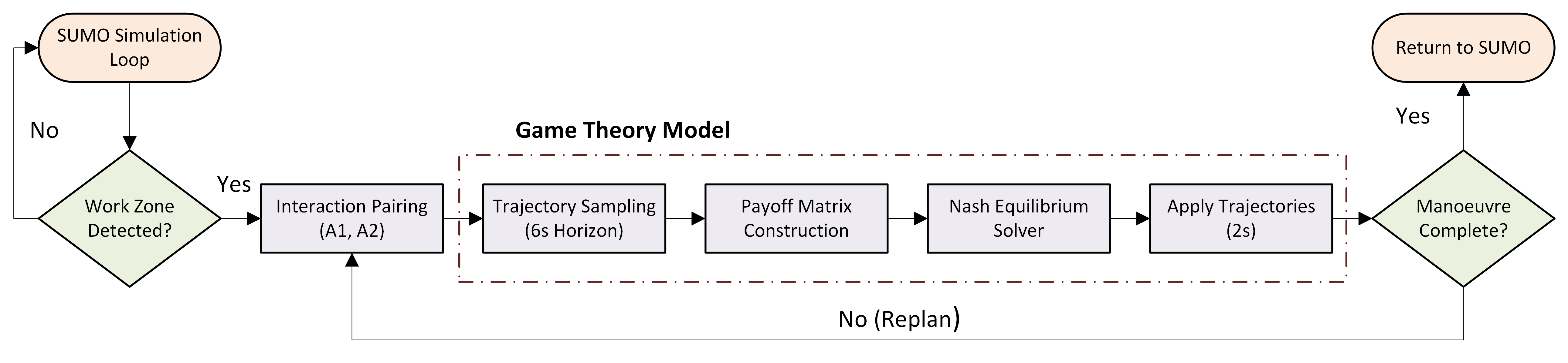}
    \caption{\footnotesize The Game-Theoretic Framework architecture.}
    \label{fig:methodology_flowchart}
\end{figure*}

The proposed framework operates as a closed-loop decision-making system integrated into the simulation environment. The operational logic of the framework is illustrated in Fig. \ref{fig:methodology_flowchart}. The process begins with identifying a mandatory lane change requirement: the ego vehicle is approaching a work zone and must transition to a target lane. This triggers the bilateral interaction pairing, which is defined as two agents, the ego and the follower in adjacent lanes, as seen in fig \ref{fig:m50_network}. Within the game-theoretic module, we solve for the Nash equilibrium of the 6-second sampled trajectories (discussed in the next subsection). Following the receding horizon principle, only the initial 2 seconds of the resulting optimal trajectory are simulated forward before replanning and solving the game again in the next iteration with the updated state of the environment. 

The following subsections detail the computational pipeline.
\subsubsection{Interaction Pairing}
At each simulation time step, the framework identifies vehicles that require a mandatory lane change due to the upcoming work zone and defines them as ego vehicles. Once an ego vehicle $v_{ego}$ is flagged, the framework defines it as agent 1 ($A_1$) and captures the current state vector for both $v_{ego}$ and its relative vehicle, which is its potential interacting partner defined as agent 2 ($A_2$). This state vector $s$ is defined as: 
$s = [x, y, v, l]$
where $x$ and $y$ represent the positions, $v$ the velocity, and $l$ represent the coordinates of the lane centrelines. These centrelines serve as reference paths for trajectory sampling.
To maintain computational efficiency and focus on the most critical interaction, the relative vehicle is defined specifically as the nearest follower in the target lane. The identification logic is based on the Euclidean distance and a directional dot product \eqref{eq:full_distance_equation}.
\begin{equation}
\begin{aligned}
d_{\text{behind}}^{*} = \min_{v \in V} \Big\{ & \|\Delta \mathbf{p}_v\|_2 \mid  \Delta \mathbf{p}_v \cdot \mathbf{u} < 0 \Big\}
\end{aligned}
\label{eq:full_distance_equation}
\end{equation}

For every vehicle $v$ in the set of candidate vehicles $V$ within the target lane, the relative displacement vector $\Delta \mathbf{p}_v$ is defined as the difference between the position of vehicle $v$ and the ego vehicle $v_{ego}$: 
\begin{equation}
   \Delta \mathbf{p}_v = [x_v - x{_{ego}}, y_v - y_{ego}]^T
   \label{eq:distance_equation}
\end{equation}

The framework then evaluates the alignment of this displacement with the ego's heading vector $\mathbf{u} = [cos \theta, sin \theta]^T $ using the dot product. A negative dot product result ($\Delta\mathbf{p}_{v} \cdot \mathbf{u} <0 $) indicates that the vehicle $v$ is positioned behind the ego vehicle. Among all vehicles that satisfy this condition, the one with the minimum Euclidean norm $||\Delta \mathbf{p}_v||_2$ is selected as the follower vehicle $v_{follower}$, which is defined as agent 2 ($A_2$). This process formalises the bilateral interaction, ensuring that the game-theoretic model accounts for the vehicle most likely to be affected by the ego vehicle's merge manoeuvrer.

\subsubsection{Game Formulation and Trajectory Sampling}
Once a pair is defined, the initial state trajectories are fed into the game-theoretic model. The game-theoretic model used in this framework follows the logic presented in \cite{sarkar2022generalized}. The model identifies a set of possible manoeuvres for both agents. Then the model samples a set of potential trajectories for both $v_{ego}$ and $v_{follower}$ on a 6-second receding horizon with a temporal resolution of $\Delta t = 0.1 s$.
The strategic interaction is modelled as a non-cooperative simultaneous dynamic game. In this formulation, each agent aims to select a trajectory that maximises its individual utility. Utility is defined as a numerical function that quantifies the desirability of a trajectory based on a balance between safety and progress. 
For each agent, a set of high-level manoeuvres is identified, including proceeding, waiting, and merging. The framework samples a set of candidate trajectories for each manoeuvrer. These trajectories are generated based on the current state and the geometric constraints of the lane centreline, ensuring that all sampled paths are kinematically feasible. A table is constructed where each entry represents the utility of a trajectory pair. The utilities for an agent $k \in [1,2]$ given a pair of trajectories $(T_{1}, T_{2})$ is defined as:
\begin{equation}
    U_{k}(T_{1}, T_{2}) = B . (w_{s}.U_{safety} + w_{p}.U_{progress} + w_{t}.U_{traffic} )
\end{equation}

Where $B=10$ is the base scaling factor. $w_{s}$, $w_{p}$, $w_{t}$ are the weights for safety, progress, and traffic, respectively. 
\begin{itemize}
    \item \textbf{Safety Weight ($w_s = 0.5$):} Prioritise collision avoidance as the primary constraint.
    \item \textbf{Progress Weight ($w_p = 0.2$):} Encourages the vehicle to complete the merge and move forward.
    \item \textbf{Traffic Weight ($w_t = 0.3$):} Penalises excessive braking.
\end{itemize}
All utilities are normalised to the range [0,1] before applying the weights.
The safety utility $U_{safety}$ quantifies collision avoidance, it is defined as:
\begin{equation}
    U_{safety} = min(1.0, \frac{max(0,d_{gap})}{d_{thr}} )
\end{equation}
Where $d_{gap}$ represents the physical gap that defines a safe distance between two trajectories. This gap is calculated as the difference between $d_{min}$, the Euclidean distance between the two agents over the trajectory horizon, and $d_{buffer}$, which accounts for the  physical dimensions of the vehicle rather than treating it as a point mass. The parameter $d_{thr}$ defines the distance beyond which the risk is considered negligible. If a collision is detected $d_{min} <= d_{buffer}$ then $U_{safety}$ = 0.
The progress utility $U_{progress}$ encourages the vehicle to move forward efficiently. To prevent an over-prioritisation of excessive speed, a logarithmic progress utility is defined as:
\begin{equation}
    U_{progress} = min(1.0, \frac{p_{prog}}{p_{thr}})
\end{equation}
Where $p_{prog}$ is defined as $\ln(path +1)$, path is the total Euclidean path length of the sampled trajectory and $p_thr$ is a normalisation factor. Using a logarithmic function ensures that the reward for progress saturates as the distance increases, prioritising steady movement over aggressive acceleration.
Finally, the traffic utility $U_{traffic}$ penalises trajectories that perform a sudden deceleration, defined as:
\begin{equation}
    U_{traffic} = min(1.0, \frac{speed_{penalty}}{speed_{thr}})
\end{equation}
Where $speed_{penalty}$ is a quadratic penalty that ensures that significant speed drops are heavily discouraged during a lane change manoeuvre, and $speed_{thr}$ is a normalisation factor.

The framework then solves for the Nash Equilibrium to identify the optimal strategy pair trajectories that represent a safe merging behaviour. A Nash Equilibrium is defined as a strategy pair in which each agent's trajectory is optimal given the trajectory chosen by the other agent.

\subsubsection{Receding Horizon Execution and Re-planning}
To ensure robustness to the dynamic traffic environment, especially in a work zone scenario, the framework adopts a receding-horizon control approach. Although the game-theoretic model generates a 6-second trajectory horizon, the agents execute only the first 2 seconds of the plan.
After this 2-second interval, the states of both agents are re-evaluated, and the re-planning phase is triggered. This closed-loop cycle of sensing, solving the Nash Equilibrium, and partial execution continues until the ego vehicle reaches the centre of the target lane, marking the lane change as completed. This iterative process enables agents to adapt to small changes in traffic flow while progressing toward a safe and successful completion of lane change.

\section{Results And Analysis}

To evaluate the performance of the proposed framework, safety metrics were analysed. A performance comparison was made between the baseline scenario, in which vehicles use the default lane change model, and the game-theoretic scenario, in which vehicles use the proposed framework perform lane changes. This comparison was made between Level 2 (L2) and Level 4 (L4) automation. We model L2 as partially automated vehicles using the C-ACC model, while L4 is fully automated using IDM model as discussed earlier. Level 3 automation was not considered, as prior studies have reported potential safety and acceptance concerns about keeping the human driver in the control loop during automated driving. These concerns suggest the transition directly from Level 2 to Level 4 \cite{gueriau2020quantifying}.
Each scenario was evaluated across 10 matched random seeds; simulations were run for 30 minutes, excluding a 5-minute warm-up to ensure robustness \cite{arvin2020safety}. The safety performance metrics assessed include the frequency of lateral conflicts per minute to quantify risks in adjacent lanes, and TTC distribution characteristics to evaluate longitudinal safety between vehicles in the same lane. In addition, the statistical significance of seeds was evaluated using a paired t-test.

\begin{figure}[tbp]
     \centering
     \begin{subfigure}[b]{0.3\textwidth}
         \centering
         \includegraphics[width=\textwidth]{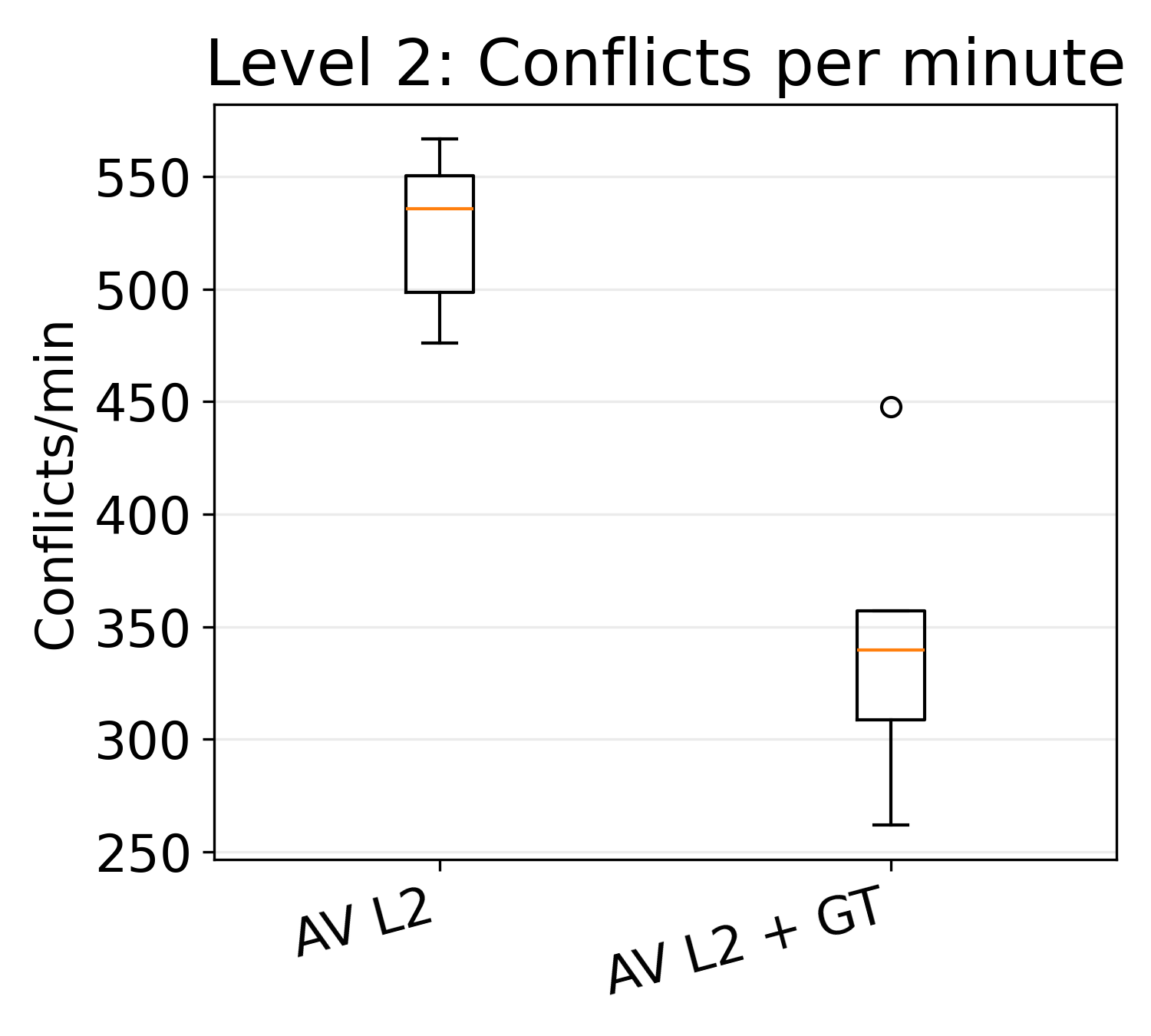}
         \caption{\footnotesize}
         \label{fig:lateral_conflicts_boxplot}
     \end{subfigure}
     \hfill
     \begin{subfigure}[b]{0.3\textwidth}
         \centering
         \includegraphics[width=\textwidth]{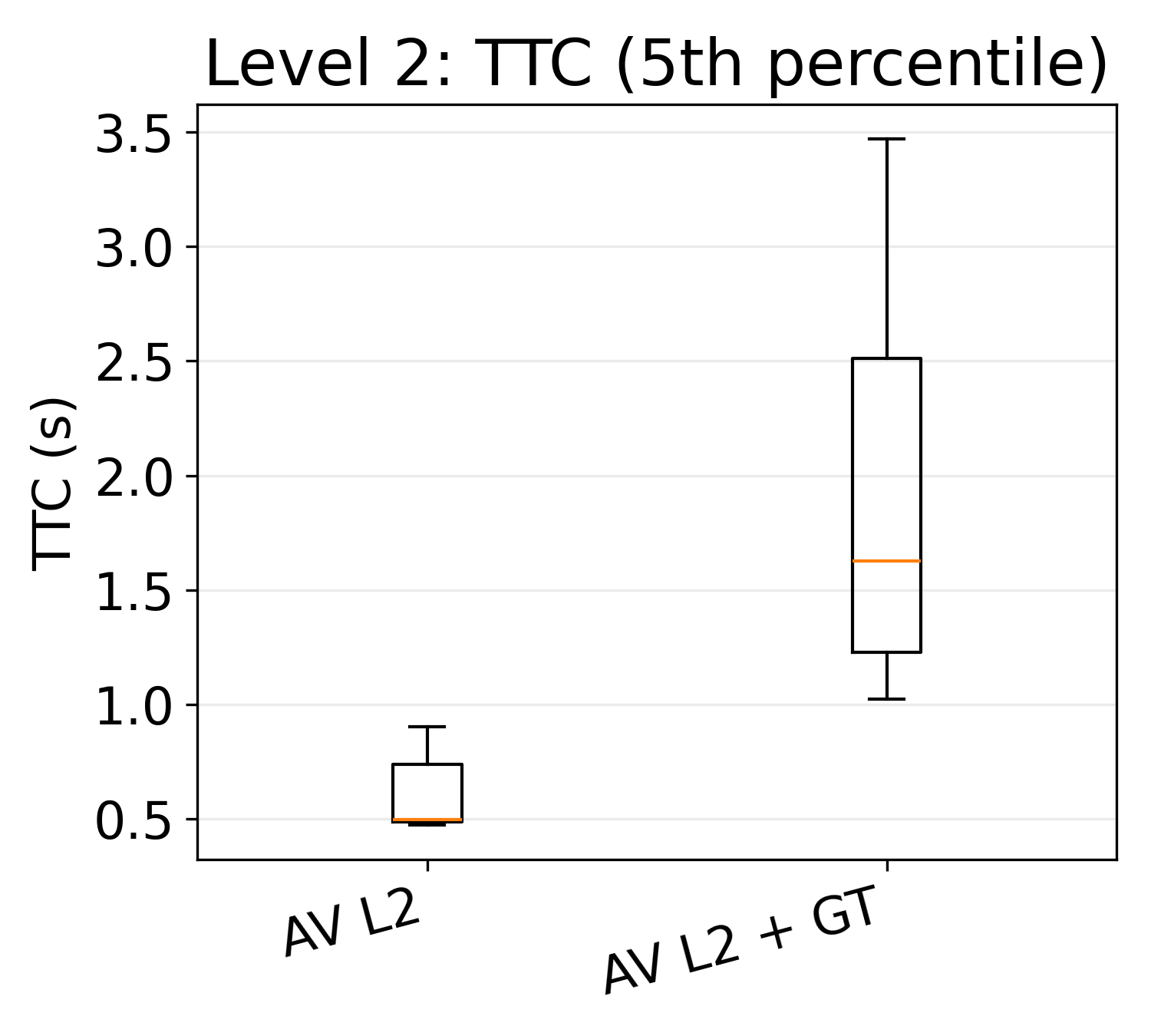}
         \caption{\footnotesize}
         \label{fig:l2_ttc_p05}
     \end{subfigure}  
     \caption{\footnotesize Fig a: Game-theoretic control reduces lateral conflict frequency under L2 automation across simulation seeds. Fig b: The 5th percentile of TTC values is higher under Game-theoretic control.}
     \label{fig:l2_conflict_findings}
\end{figure}

In L2 automation, a reduction of approximately 35\% is seen for the number of lateral conflict frequency between vehicles in a work zone environment. Figure \ref{fig:lateral_conflicts_boxplot} shows the average conflict rate among seeds for Level 2 automation. The baseline L2 scenario exhibits a high conflict rate of $\approx$ 525 conflicts per min on average, whereas the proposed game-theoretic model reduces this to $\approx$ 343 conflicts per min. The proposed framework reduced the number of times vehicles encountered a lateral near miss with a TTC or a time gap less than the defined safety thresholds (2 sec for TTC and 1 m for time gap) when performing a lane change before the lane closure.
The paired t-test confirms statistical significance (p = 0.0048), indicating that the reduction in conflicts is consistent between seeds. The 95\% confidence intervals show limited overlap across seeds, indicating that the improvement is not due to random seed variation; the effect size exceeds the noise at seed-level, further supporting robustness.
For better understanding, we analysed the fifth percentile of TTC to analyse the worst 5\% interactions. Figure \ref{fig:l2_ttc_p05} shows that the lower tail TTC,5th percentile, increased from $\approx$0.6 s to $\approx$1.7 s, indicating an improvement in minimum safety margins.

\begin{figure}[tbp]
     \centering
     \begin{subfigure}[b]{0.4\textwidth}
         \centering
         \includegraphics[width=\textwidth]{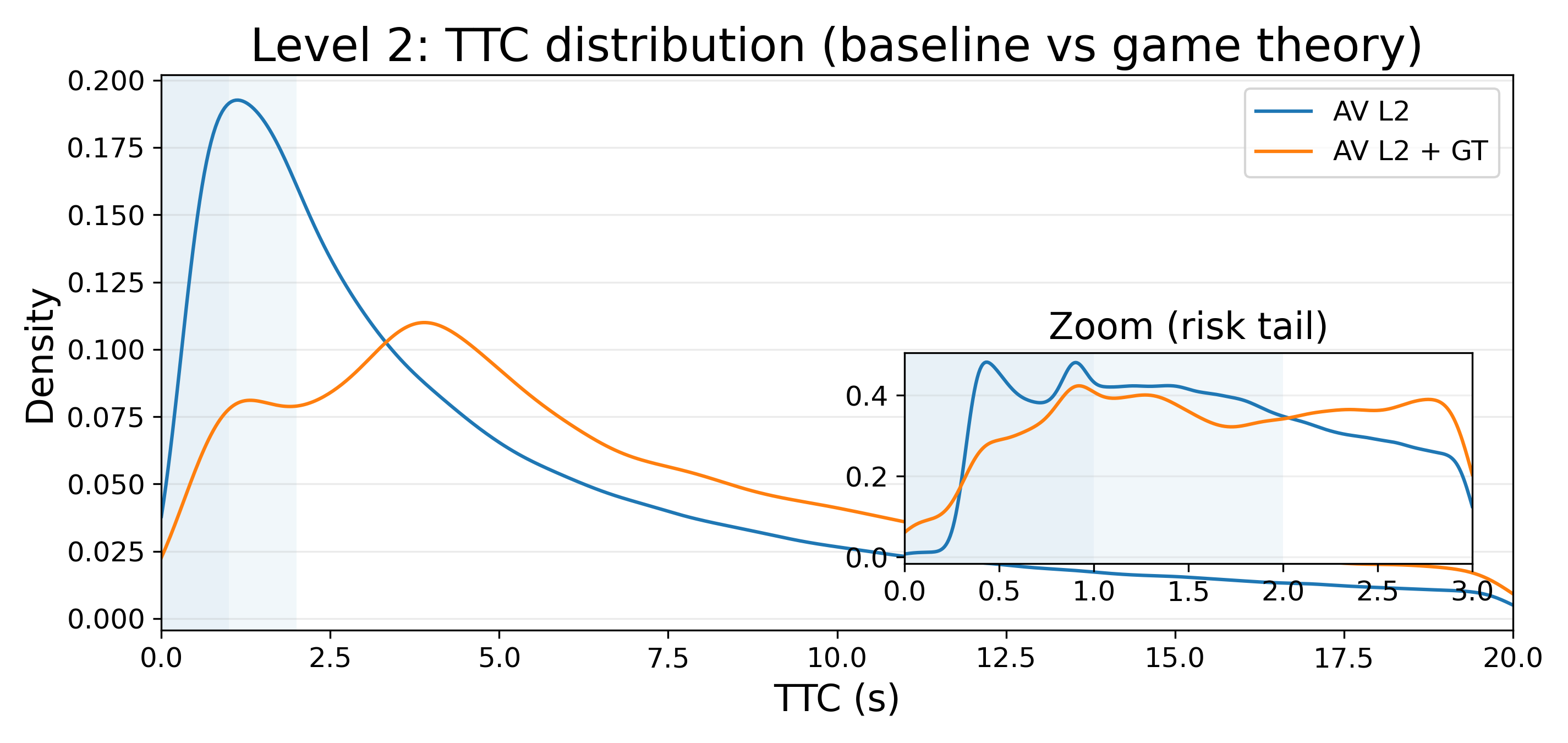}
         \caption{\footnotesize TTC PDF distribution}
         \label{fig:long_TTC_pdf}
     \end{subfigure}
     \begin{subfigure}[b]{0.4\textwidth}
         \centering
         \includegraphics[width=\textwidth]{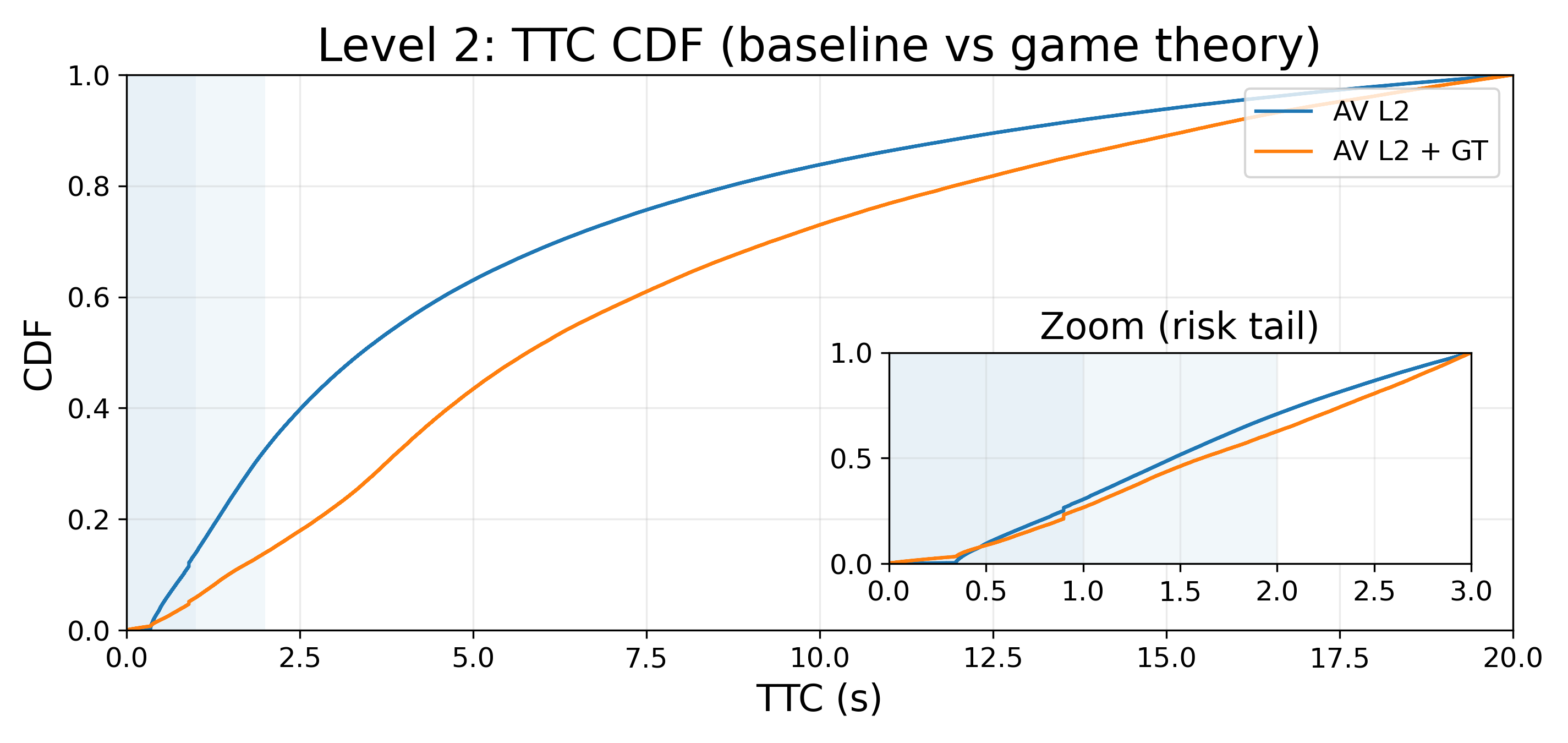}
         \caption{\footnotesize TTC CDF distribution}
         \label{fig:long_TTC_cdf}
     \end{subfigure}
     \hfil
     \caption{\footnotesize Under L2, the proposed framework shifts the TTC distribution toward higher values, reducing the probability of high-risk interactions ($TTC < 2$ s) compared to the baseline scenario. Risk regions are defined as high risk ($< 2$ s), moderate risk (2–3 s), and safe ($> 3$ s).}
     \label{fig:l2_TTC_distributions}
\end{figure}

Figures \ref{fig:long_TTC_pdf} and \ref{fig:long_TTC_cdf} illustrate the TTC distribution. The TTC is calculated longitudinally between a following vehicle and its leader in the same lane, upstream of the work zone lane closure as defined in equation \ref{eq:ttc_calc}.
\begin{equation}
    \label{eq:ttc_calc}
    TTC_i(t) = \frac{x_{i-1}(t) - x_i(t) - L_{i-1}}{v_i(t) - v_{i-1}(t)} \quad \text{if } v_i(t) > v_{i-1}(t)
\end{equation}
The Probability Density Function (PDF) is estimated using kernel density estimation and normalised so that the total probability integrates to one. The inset panel provides a magnified view of the safety critical region $(0-3 s)$ with independently scaled axes to make the tail risk visible.
The PDF shows a low concentration of TTC values $(< 2 s)$ in the baseline scenario, indicating more high-risk interactions. On the other hand, the game-theoretic model shifts the probability mass to higher TTC values, indicating lower-risk interactions.
The Cumulative Density Function (CDF) confirms this shift. In the high-risk region $(< 2 s)$, the probability of the baseline exceeds that predicted by the game-theoretic model. In the safety-critical region, the baseline CDF accumulates probability faster than the game-theoretic model, indicating a higher likelihood of unsafe interactions.
To evaluate more safety-critical thresholds, we calculated the number of events with TTC $( < 2 s)$ between TTC samples for each scenario among seeds, and the paired t-test confirms statistical significance (p = 0.008), indicating that the proposed model decreases high-risk lateral interactions. Table \ref{tab:paired_ttest_results} shows the summary of the paired t-tests.

On the other hand, in L4 automation, improvements are statistically non-significant. The baseline L4 scenario already enforces low conflict rates and high TTC values by maintaining a safe gap. The baseline L4 scenario has $\approx$ 288 conflicts per minute and the game theoretic model $\approx$ 286 conflicts per minute. the difference is minimal and statistically not significant $(p \approx 0.82)$. This indicates that fully automated vehicles already maintain stable interaction behaviour even in a work zone, leaving limited room for improvement.

\begin{figure}[tbp]
     \centering
     \begin{subfigure}[b]{0.4\textwidth}
         \centering
         \includegraphics[width=\textwidth]{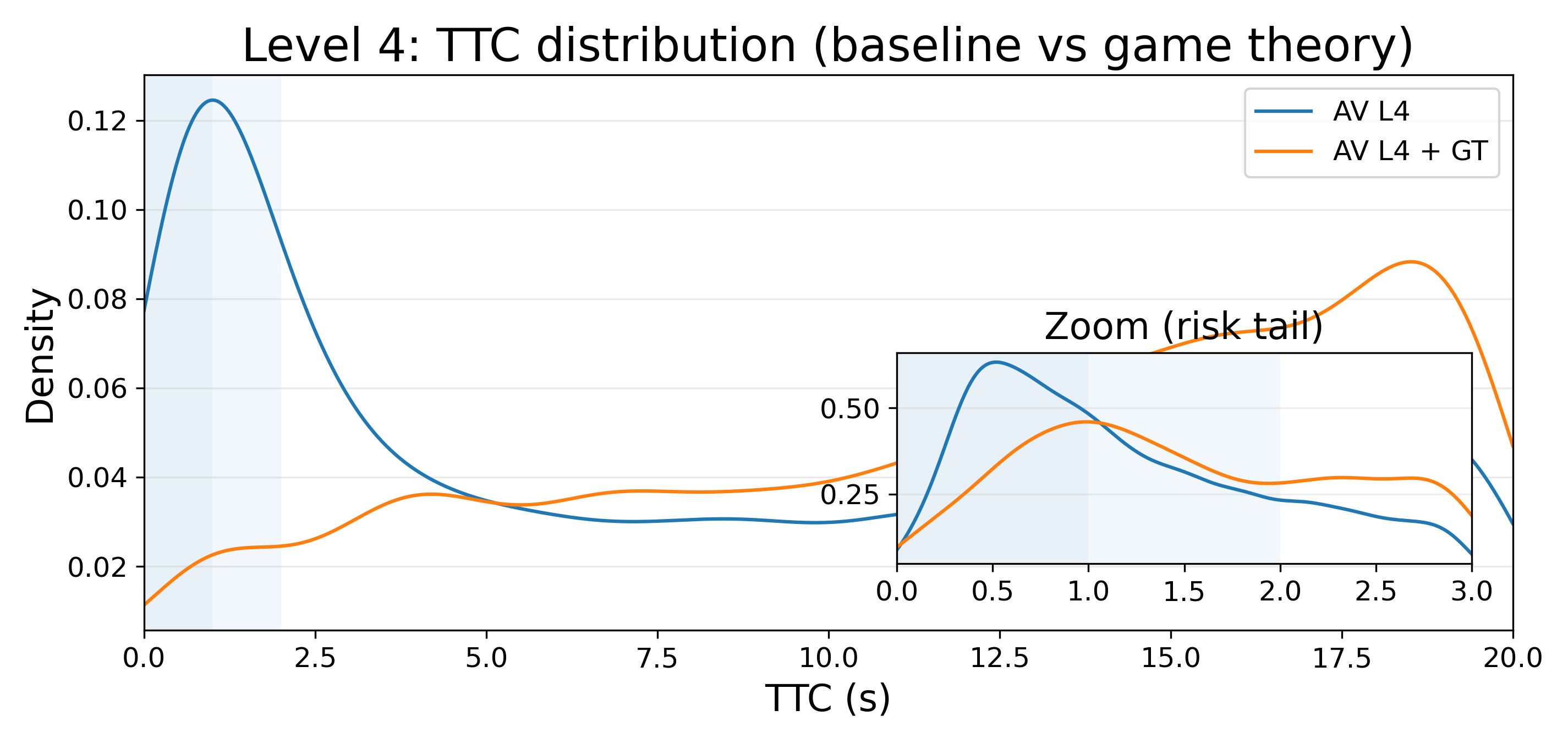}
         \caption{\footnotesize TTC PDF distribution}
         \label{fig:l4_TTC_pdf}
     \end{subfigure}
     \begin{subfigure}[b]{0.4\textwidth}
         \centering
         \includegraphics[width=\textwidth]{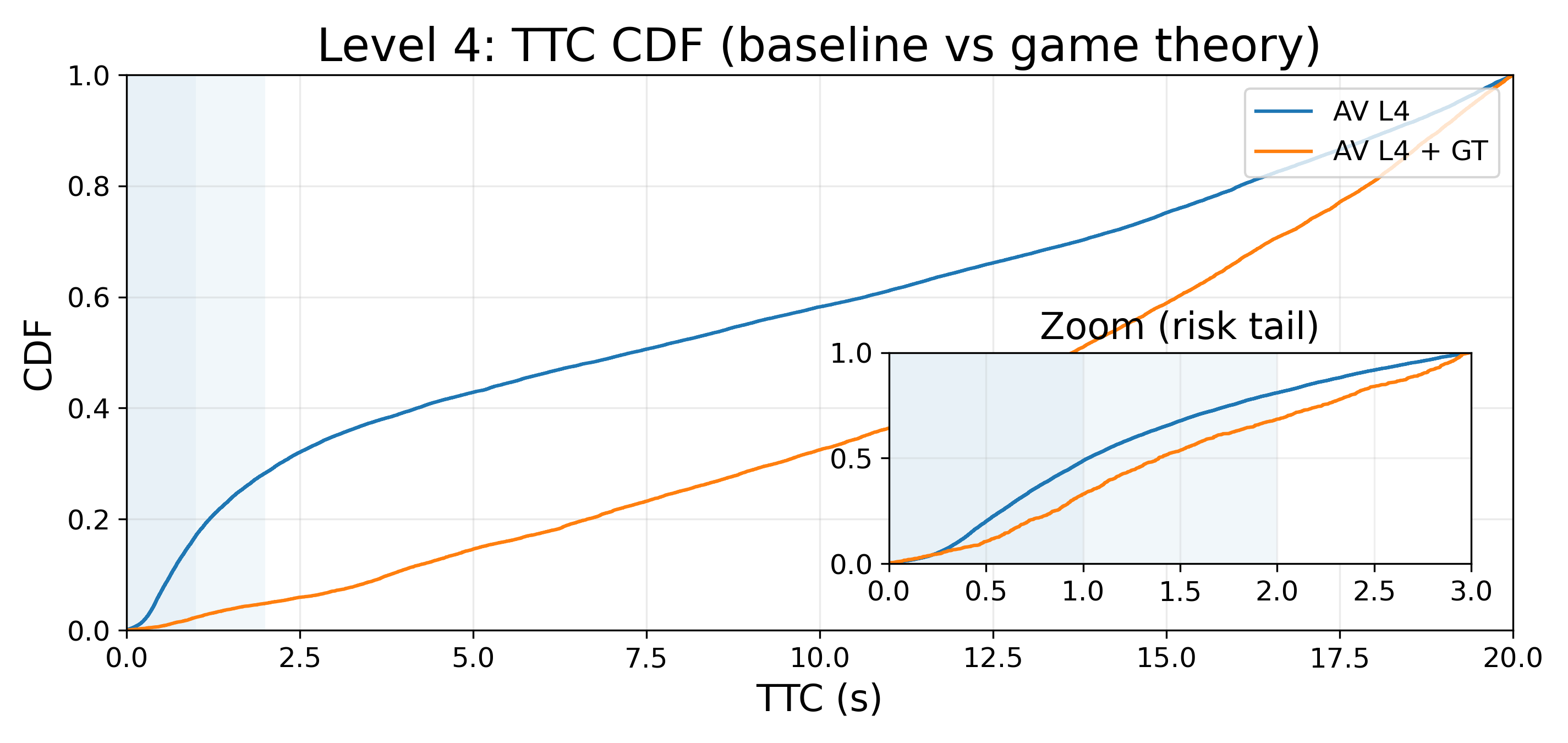}
         \caption{\footnotesize TTC CDF distribution}
         \label{fig:l4_TTC_cdf}
     \end{subfigure}
     \hfil
     \caption{\footnotesize Under L4, the proposed framework produces a slight rightward shift in the TTC distribution. Risk regions are defined as high risk ($< 2$ s), moderate risk (2–3 s), and safe ($> 3$ s).}
     \label{fig:l4_TTC_distributions}
\end{figure}

When analysing the CDF and PDF of the L4 baseline scenario against the L4 game theoretic scenario, as seen in figures \ref{fig:l4_TTC_pdf} and \ref{fig:l4_TTC_cdf}, we observe a rightward shift in TTC distribution in the game theoretic model and a lower probability accumulation in the risk region $( < 2 s)$. However, when we calculated the number of events with TTC $( < 2 s)$ for each scenario, the paired t-test showed a non-statistically significant value $p = 0.27$, which indicates that level 4 automation may already operate safely, and the game theoretic model just forces the shift to wider gaps between vehicles in lateral interaction.

\begin{table}[tbp]
\centering
\setlength{\tabcolsep}{3pt}
\caption{Paired t-test results across matched simulation seeds (N = 10).}
\label{tab:paired_ttest_results}
\begin{tabular}{lcccc}
\hline
\textbf{Automation} & \textbf{Metric} & \textbf{Baseline Mean} & \textbf{Game Mean} & \textbf{$p$-value} \\
\hline
L2 & Conflicts/min & 525.5 & 342.9 & 0.005 \\
L2 & TTC p05 (s) & 0.62 & 1.97 & 0.048 \\
L2 & TTC $<$ 2 s (\%) & 27.97 & 6.18 & 0.0085 \\

L4 & Conflicts/min & 288.4 & 286.1 & 0.8209 \\
L4 & TTC p05 (s) & 11.51 & 16.05 & 0.15 \\
L4 & TTC $<$ 2 s (\%) & 2.76 & 0.31 & 0.2746 \\
\hline
\end{tabular}
\end{table}

\section{Conclusion}

This study proposed a game-theoretic framework for lane change manoeuvres in work zone environments and evaluated its impact on safety under Level 2 (L2) and Level 4 (L4) automation, representing partial and full automation, respectively. To assess its feasibility, the framework was integrated in a calibrated microscopic traffic simulation environment and evaluated using surrogate safety measures across 10 matched random seeds. The results indicate that in L2 automation, the proposed model significantly improves safety performance. The frequency of lateral conflicts was reduced by approximately 35\%, and statistically significant reductions were observed in TTC events. These findings demonstrate that incorporating a strategic game-theoretic model for manoeuvres such as lane changes mitigates high-risk conflicts in work-zone environments. In contrast, under L4 automation, improvements were statistically non-significant. This indicates that the baseline L4 vehicles already maintained low conflict rates and high TTC values.
Future work will extend the framework to mixed-traffic conditions, including varying market penetration rates in the game-theoretic model, to investigate scalability. In addition, extending the framework to include the leader vehicle in the game and simulating actual communication would improve safety and allow modelling of gap-acceptance behaviour among the ego, leader, and follower vehicles\cite{nour2025assessing}\cite{nour2025integrating}. These extensions will enhance the realism and scalability of the proposed framework for connected and automated vehicle lane-change decision-making.

\section{Acknowledgement}
The authors thank the National Cybersecurity Consortium for their support of this work

\bibliographystyle{IEEEtran}
\bibliography{ref}

\end{document}